\documentclass[12pt,preprint]{aastex}

\slugcomment{Accepted for publication in Astophys. J.}

\shorttitle{}
\shortauthors{}

\begin{document}



\title{The effects of a $\kappa$-distribution in the heliosheath on
  the global heliosphere and ENA flux at 1 AU}


\author{J. Heerikhuisen, N.V. Pogorelov, V. Florinski, G.P. Zank and
  J. A. le Roux}
\affil{Institute of Geophysics and Planetary Physics, University of California,
    Riverside, CA 92521}
\email{jacobh@ucr.edu}
\email{nikolai.pogorelov@ucr.edu}
\email{vflorins@ucr.edu}
\email{zank@ucr.edu}
\email{jakobus.leroux@ucr.edu}

\begin{abstract}

By the end of 2008 (approximately one year, at the time of writing),
the NASA SMall EXplorer (SMEX) mission IBEX (Interstellar Boundary
Explorer) will begin to return data on the flux of energetic neutral
atoms (ENA's) observed from an eccentric Earth orbit. This data will
provide information about the inner heliosheath (the region of
post-shock solar wind) where ENA's are born through charge-exchange
between interstellar neutral atoms and plasma protons. However, the
observed flux will be a function of the heliosheath thickness, the
shape of the proton distribution function, the bulk plasma flow, and
loss mechanisms acting on ENA's traveling to the detector.  As such,
ENA fluxes obtained by IBEX can be used to better parametrize global
models which can then provide improved quantitative data on the shape
and plasma characteristics of the heliosphere.  In a recent letter
\citep{HPZandF07a}, we explored the relationship between various
geometries of the global heliosphere and the corresponding ENA all-sky
maps. There we concentrated on energies close to the thermal core of
the heliosheath distribution (200 eV), which allowed us to assume a
simple Maxwellian profile for heliosheath protons. In this paper we
investigate ENA fluxes at higher energies (IBEX detects ENA's up to 6
keV), by assuming that the heliosheath proton distribution can be
approximated by a $\kappa$-distribution. The choice of the $\kappa$
parameter derives from observational data of the solar wind (SW).  We
will look at all-sky ENA maps within the IBEX energy range, as well as
ENA energy spectra in several directions. We find that the use of
$\kappa$ gives rise to greatly increased ENA fluxes above 1 keV, while
medium energy fluxes are somewhat reduced. We show how IBEX data can
be used to estimate the spectral slope in the heliosheath, and that
the use of $\kappa$ reduces the differences between ENA maps at
different energies. We also investigate the effect introducing a
$\kappa$-distribution has on the global interaction between the SW and
the local interstellar medium (LISM), and find that there is generally
an increase in energy transport from the heliosphere into the LISM,
due to the modified profile of ENA's energies. This results in a
termination shock that moves out by 4 AU, a heliopause that moves in
by 9 AU and a bow shock 25 AU farther out, in the nose direction.

\end{abstract}



\section{Introduction}

With the crossing of the termination shock (TS) by the {\it Voyager} 1
and 2 spacecraft \citep{BNALCSandMcD05,DKRHAGHandL05,SCMcDHLandW05},
the post-shock solar wind (SW) region, known as the inner heliosheath
\citep{Zank99}, has become an area of increased interest
\citep{igpp_conf6}. Despite its non-functioning plasma instrument,
      {\it Voyager} 1 has provided important data on the flow,
      energetic particle, and magnetic field orientation in the
      heliosheath, much of which is poorly understood. Now that {\it
        Voyager} 2 has crossed the TS at 84 astronomical units (AU),
      new data will further increase our understanding of the outer
      reaches of the heliosphere.

Although \emph{in situ} measurements by the {\it Voyager} spacecrafts
are immensely valuable, they do not provide much information about the
global structure of the heliosphere-interstellar medium interaction
region. The Interstellar Boundary Explorer \citep[{\it
    IBEX},][]{McComas_IGPP04,McComas_IGPP06} will try to infer global
heliospheric structure by surveying the sky in energetic neutral atoms
(ENA's) from Earth orbit. ENA's are created in the heliosheath after a
neutral atom from the local interstellar medium (LISM)
charge-exchanges with a plasma proton. The new neutral atom (generally
hydrogen) is born from the proton distribution, and, as such, reflects
the characteristic plasma conditions at the point of creation. ENA's
propagate virtually ballistically (particularly ENA hydrogen), subject
only to the sun's gravity and radiation pressure. {\it IBEX} will
directly detect ENA's and create all-sky maps at a variety of energies
between 10 eV and 6 keV at the rate of one complete map every six
months.

The challenge to both data analysts and theorists is how to interpret
the ENA flux measurements made by the IBEX-Lo (10 eV -- 2 keV) and
IBEX-Hi (300 eV -- 6 keV) instruments. The ENA flux at a given energy
will be a function of the properties of the heliosheath along a
particular line of sight. As shown in \cite{HPZandF07a}, this
includes plasma and neutral number densities, plasma flow speed and
direction, plasma temperature, and distance to the heliopause
(heliosheath thickness). However, that analysis was limited
to energies close the thermal core of the heliosheath distribution,
since we did not incorporate high energy tails in the ENA parent
population due to either pick-up ions, or energetic protons
accelerated by other mechanisms.

Recently, \cite{Prested_kappa08} used a $\kappa$-distribution for the
ENA parent population to obtain ENA maps. The advantage of using this
distribution, as opposed to a Maxwellian, is that it has a power-law
tail, and is therefore capable of producing ENA's at suprathermal
energies. However, the focus in that paper was on the {\it IBEX}
instrument's response to ENA fluxes, and feed-back of ENA's on the
global solution was not considered.

In this paper we seek to extend the investigations of
\cite{HPZandF07a} to higher energies by adopting a
$\kappa$-distribution for heliosheath protons, using an approach
similar to \cite{Prested_kappa08}. The suggestion that the supersonic
SW should be described by a $\kappa$-distribution rather than a
Maxwellian has a long history \citep{GABFZPSandH81,SandT91}. More
recently, with the measurement of PUI's by {\it Ulysses}
\citep{GFandL05,FandG06}, it became apparent that the PUI distribution
merged cleanly into the solar wind distribution, yielding an extended
energetic tail. This was carried further by \cite{Mewaldt_etal01} who
constructed an extended supersonic SW proton spectrum showing that a
high energy tail emerged smoothly from the clearly identifiable low
energy solar wind particles. The results of \cite{Mewaldt_etal01}
showed that not only did a continuous power law tail emerge from the
thermal distribution, but this tail merged naturally into higher
energies associated with (low energy) anomalous cosmic rays (ACR's)
\citep{DKRHAGHandL05}. The Voyager LECP data obtained in the
heliosheath indicates that a power law distribution at thermal
energies is maintained, but of course we have no means to show that a
tail emerges smoothly from the shocked SW plasma. Nonetheless, we do
not expect an abrupt departure from the supersonic SW particle
distribution characteristics in that its overall ``smoothness'' should
be preserved.

We use a self-consistently coupled MHD-plasma/kinetic-neutral code to
compute a steady-state heliosphere with a $\kappa$-distribution in the
SW, and investigate ENA fluxes at 1 AU, looking in particular for
signatures which can be related to the heliospheric structure. We
begin, however, by investigating the effects of assuming such a
distribution on the supersonic and subsonic SW and, due to the
non-local coupling mediated by charge-exchanging neutrals, the global
heliosphere.

\section{The heliosphere with $\kappa$ heliosheath}

\begin{figure}[ht!]
\epsscale{1.0}
\plotone{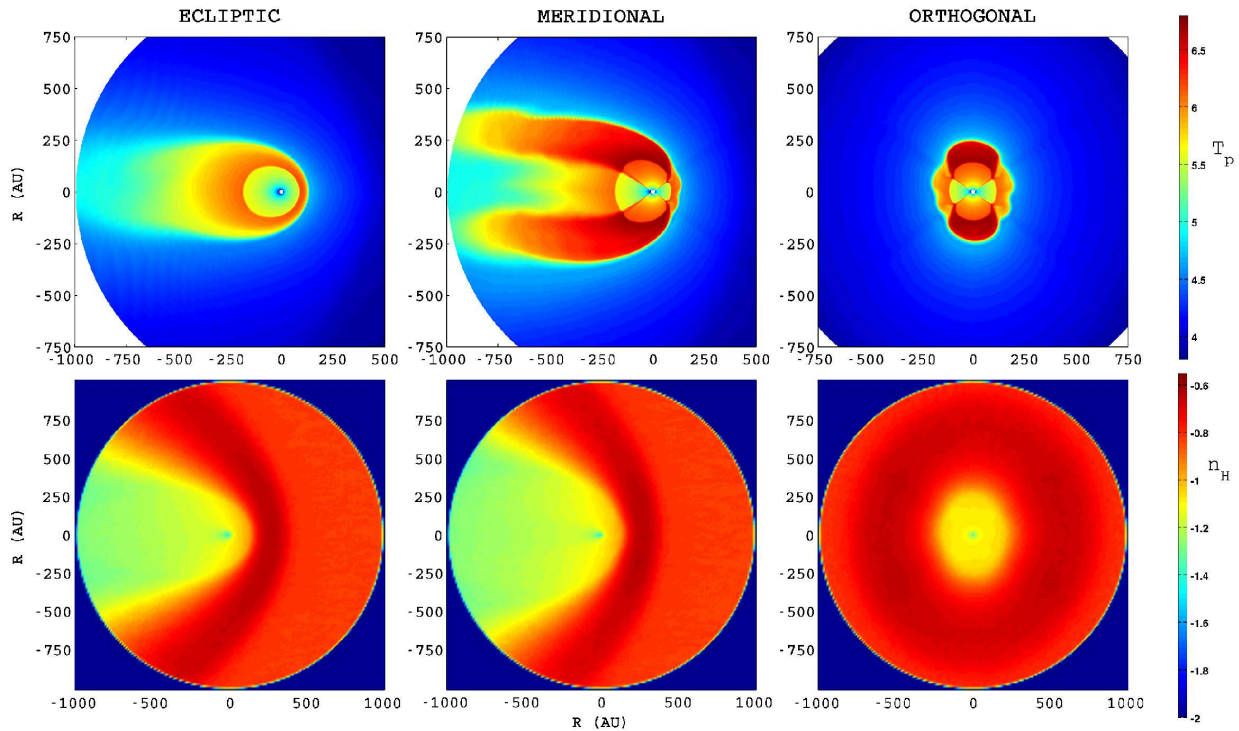}
\caption{Global heliospheric solution with the boundary conditions
  described in Table \ref{table:3D_bc}. The three columns represent
  cuts of the heliosphere through the Sun along the ecliptic plane
  (left), meridional plane (middle), and the plane orthogonal to the
  LISM flow vector. The top row is a log$_{10}$ plot of plasma
  temperature in K, while the bottom row is a log$_{10}$ plot of
  neutral density in cm$^{-3}$. Distances are in astronomical units
  (AU). Note how the streams of high speed SW over the poles generate
  hotter subsonic SW in the heliosheath \citep{PandZ96,PandZ97}. This
  high speed wind also symmetrizes the heliopause near the Sun,
  despite the presence of LISM magnetic field which generally acts to
  asymmetrize the heliosphere \citep{PZandO04,OSandL06}, although
  noticeably less so when neutrals are taken into account
  \citep{PandZ06b,PSFandZ07}. The build-up of neutral hydrogen just
  outside the heliopause, known as the ``hydrogen wall'', can be
  clearly seen in the lower plots.}
\label{fig:HPFZandL07_global}
\end{figure}

At around 100 astronomical units (AU) the supersonic SW flow
encounters the termination shock (TS), whereupon it becomes subsonic
and heated. The hot subsonic SW fills the inner heliosheath and
heliotail (these features are visible in the computed plasma
distributions shown in Figure \ref{fig:HPFZandL07_global}). At the
same time, the solar system is thought to travel supersonically
through the partially ionized plasma of the LISM. As a result, a bow
shock forms upstream of the heliosphere, and a tangential
discontinuity, known as the heliopause (HP), separates the shocked
solar and LISM plasmas. Interstellar neutral gas (primarily hydrogen)
is weakly coupled to the plasma through charge-exchange, but readily
traverses the heliopause (with a filtration ratio of about 45\%) and
may be detected near Earth at a range of energies that correspond to
the creation site of the neutral H, ranging from the LISM to the hot
heliosheath, to the fast solar wind.

To determine the flux of neutral atoms at 1 AU, we use a steady-state
solution obtained from the 3D heliospheric model based on the 3D MHD
code of \cite{PZandO06} and a 3D version of the kinetic neutral
hydrogen code of \cite{HFandZ06}. The first self-consistently coupled
3D application of this code appears in \cite{PHandZ08}.  A
steady-state is reached by iteratively running the coupled plasma and
neutral codes until successive iterations converge.  Although several
plasma-only models of the heliosphere are still in use, it is now
recognized that including neutral atoms in a global model is critical
to obtaining the correct location and shape of the termination shock
and heliopause, as well as determining the right temperature of the
heliosheath, since interstellar neutrals contribute to significant
cooling and heating of the inner and outer heliosheath respectively
\citep{PSFandZ07}. We also note that inter-particle collisions do not
significantly alter the neutral distribution and that charge-exchange
mean free paths are of the order of the size of the heliosphere, so
that neutral atoms should ideally be modelled kinetically, with
charge-exchange coupling the neutral and charged populations
\citep{BandM93,AandI05,HFandZ06}.

\begin{table}[ht!]
\begin{center}
\begin{tabular}{|l|c|cc|}
\hline
Parameter & Interstellar & \multicolumn{2}{c|}{1 AU} \\
& & Low Speed & High Speed \\
\hline
$U$ (km/s)           & 26.4   & 400    & 800 \\
$T$ (K)              & 6527   & $10^5$ & 2.6$\times 10^5$ \\
$n_p$ (cm$^{-3}$)     & 0.05   & 7      & 2.6  \\
$n_H$ (cm$^{-3}$)     & 0.15   & 0      & 0 \\
$|B|$ ($\mu$G)       & 1.5    & 37.5 ($B_r$)  &37.5 ($B_r$)\\
$\phi_B$ ($^\circ$)   & 90     &        & \\
$\theta_B$ ($^\circ$) & 60     &        & \\
\hline
\end{tabular}
\end{center}
\caption{Boundary conditions for the 3D heliospheric model considered
  here. We use a spherical coordinate system, where $\phi$ is the
  angle in the ecliptic plane around from the meridional plane and
  $\theta$ is the angle above the ecliptic plane. The solar rotation
  axis is assumed orthogonal to the ecliptic plane. The SW is assumed
  to change from a slow wind to a high speed wind at 35 degrees above
  the ecliptic plane, as suggested by Ulysses observations
  \citep{McComas_etal00_Ulysses} of the SW during solar minimum.}
\label{table:3D_bc}
\end{table}

\begin{figure}[ht!]
\epsscale{1.0}
\plotone{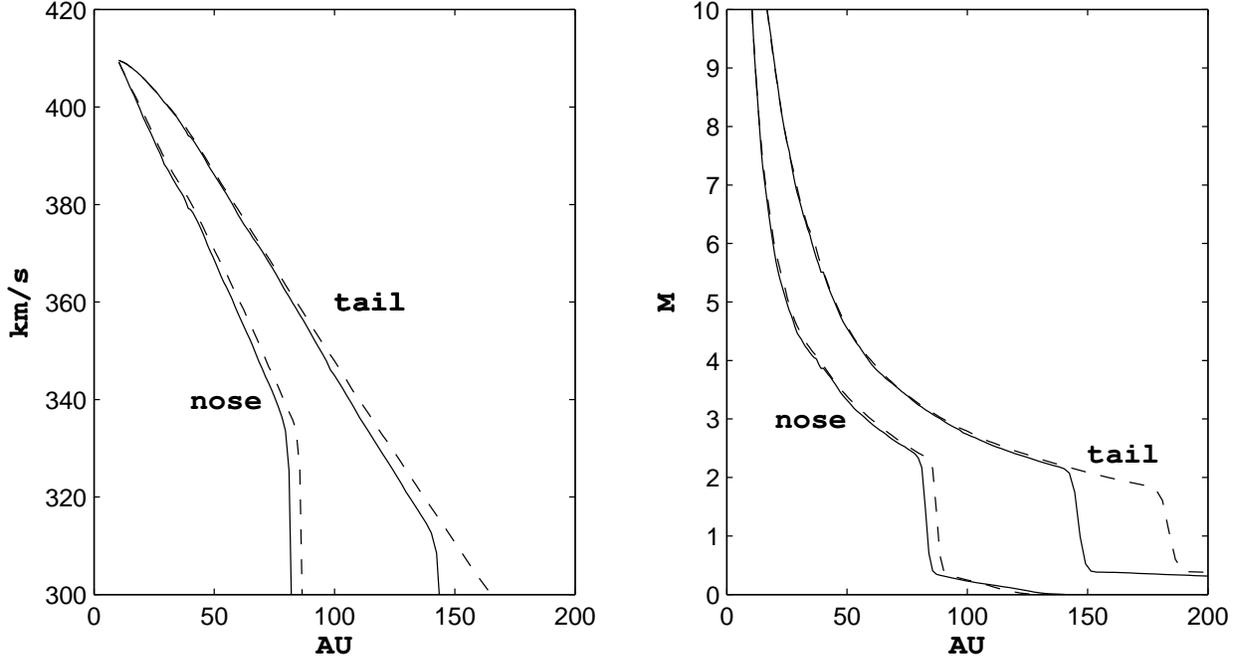}
\caption{
The solar wind bulk speed (left), and the corresponding Mach
  number as computed from (\ref{eq:mach}). Here we have plotted
  profiles in both the LISM upwind (nose) and downwind (tail) directions for a model
  using Maxwellian (solid) and $\kappa$ (dashed) distributions for the
  solar wind. In our calculation the TS has a Mach number of about 2.3
  in the nose direction, and around 2 in the tail. Note also the
  asymmetry in the solar wind speed from nose to tail, due to the
  reduced charge-exchange rate in the tail. The SW speed at the inner
  boundary, located at $r = 10$ AU, is slightly higher than indicated
  in Table \ref{table:3D_bc} due to the thermal acceleration of the SW
close to 1 AU.}
\label{fig:Mach_speed}
\end{figure}

Our model treats the ion population as a single fluid whose total
pressure is the sum of the pressure contribution from electrons,
thermal ions (SW or LISM), and PUI's. Because the pick-up of
interstellar neutral H yields a PUI population co-moving with the bulk
SW flow, a single fluid model captures exactly the energetics and
dynamics of the combined SW/PUI plasma. The only assumption that is
needed is for the value of the adiabatic index ($\gamma = 2$
corresponds to no scattering of the PUI distribution, $\gamma = 5/3$
corresponds to scattering of the PUI's onto a shell distribution) --
see, for example, \cite{KSZandP96} or section 4.1 of
\cite{Zank99}. The pick-up of ions and the creation of new H-atoms is
included self-consistently through source integrals in the plasma
momentum and energy equations \citep{Holzer72,PZandW95}. The pick-up
of interstellar neutrals and the creation of PUI's in the supersonic
SW removes energy and momentum from the SW since the newborn ions are
accelerated in the SW motional electric field to co-move with the SW
flow. The fast neutrals created in the supersonic SW propagate
radially outward, typically experiencing charge-exchange in the
LISM. Pick-up of neutrals in the SW therefore decelerates the flow,
and since a population of PUI's with thermal velocities comparable to
the bulk SW speed ($\sim 1$ keV energies) is created, the {\it total}
pressure/temperature in the one-fluid model begins to increase with
increasing heliocentric radius. Of course, the thermal SW ions
experience no heating other than due to enhanced dissipation
associated with excitation of turbulence by the pick-up process
\citep{WZandM95,ZMandS96}. These effects are all captured by the
self-consistent coupling of plasma, via a one-fluid plasma model, and
neutral H, and the plasma pressure and velocity respond directly to
the distribution of neutral H throughout the heliosphere. Finally, as
neutral H drifts through the heliosphere from the upwind to downwind,
neutral H is depleted leading to less pick-up towards the heliotail
region. This results in a (relatively weak) upwind-downwind asymmetry
in the SW plasma flow velocity 
(see Figure \ref{fig:Mach_speed}, below)
and the one-fluid (i.e. PUI's) pressure/temperature. It should
be noted that these results are independent of the specific form of
the plasma ion (thermal and PUI) distribution function, as long as it
is assumed isotropic. Only in computing the specific source term for
both the plasma and neutral equations does the detailed distribution
become important, and then primarily for the neutral distribution
(since new-born PUI's are always accelerated by the motional electric
field to co-move with the SW flow).

What we have just described is the heating/pressurization of a single
fluid SW due to charge-exchange with interstellar Hydrogen. Our
$\kappa$-distribution approach tries to improve on this by using a
distribution with core and tail features to approximate the core SW,
suprathermal ion, and PUI distributions respectively. Of course in
reality the solar wind is much better described by separate
distributions. In fact, a drawback of our approach is that the value
of $\kappa$ we use fixes the ratio between the core and tail number
densities so that one cannot change independently characteristics of
the core without making self-similar change to the wings of the
$\kappa$-distribution. In particular, this manifests itself in the
radial temperature profile of the solar wind. Observations by
\cite{RPLandB95} suggest that the core SW does not cool
adiabatically, but instead appears to be heated. New-born PUIs form an
unstable ring-beam distribution which excites Alfv\'en waves that then
scatter the PUIs onto a bispherical distribution. The power in the
excited waves can be computed geometrically as the difference in the
energy between the an energy conserving shell distribution for PUIs
and a bispherical distribution for PUIs \citep{WandZ94} or directly
from quasi-linear theory \citep{LandI87}. To explain the heating
observed by \cite{RPLandB95}, \cite{WZandM95} suggested that the
dissipation of the PUI excited waves could account for the heating,
but it was only with the development of a transport model for magnetic
field fluctuations and their turbulent dissipation (which leads to
heating of the plasma) that the PUI excited fluctuations be properly
accounted for \citep{ZMandS96}. Since the dissipation of magnetic
fluctuation power is strengthened in the outer heliosphere by PUI
excited fluctuations, this leads to a corresponding heating of the
solar wind plasma in the outer heliosphere. \cite{MZSandO99} applied
the turbulence transport model of \cite{ZMandS96} to show explicitly
that PUI enhanced turbulent dissipation of magnetic field fluctuations
could account for the observed solar wind plasma heating, a result
that was examined in considerably more detail by \cite{SMZNOandR01}
\citep[see also][]{CFandL03,SIMandR06}. The dissipation of magnetic
energy affects only the solar wind core, heating it, but leaves the
suprathermal and PUI population unchanged energetically. Within a
single fluid description, both the core and tail components of the
distribution broaden simultaneously, and we cannot alter the ratio of
energization between these components, as would be required if we were
to account for turbulent dissipation of magnetic fluctuation energy
into the solar wind plasma. Nonetheless, the total dynamics of the
system, including charge exchange levels, is preserved but the
detailed energy allotment between the core SW and PUI's is fixed by
the choice of the $\kappa$ parameter.

Figure \ref{fig:HPFZandL07_global} shows cuts of the heliosphere in
three planes for the plasma temperature and neutral hydrogen
density. These results were obtained using our 3D
MHD-plasma/kinetic-neutral model, where we assumed a
$\kappa$-distribution for protons in the heliosheath with $\kappa =
1.63$. The SW and LISM boundary conditions used in this calculation
are summarized in Table \ref{table:3D_bc}.  As described above,
  the pick-up process for our single ion fluid approach results in
  solar wind properties expected from observational data --
  i.e. increased pressure and decreased speed at larger radial
  distances. To demonstrate this using our code, Figure
  \ref{fig:Mach_speed} shows profiles of the bulk speed of the SW, and
  the fast magnetosonic Mach number given by
\begin{equation}\label{eq:mach}
M = 2 u_r / \left(\sqrt{c_s^2 + \frac{B^2}{4\pi\rho} +
  \frac{|B_r|c_s}{\sqrt{\pi\rho}}} + \sqrt{c_s^2 + \frac{B^2}{4\pi\rho} -
            \frac{|B_r|c_s}{\sqrt{\pi\rho}}} \right) \; ,
\end{equation}
where $\rho$, $P$ and $c_s^2 = \gamma P/\rho$ are the plasma density,
pressure and sound speed respectively. The adiabatic index $\gamma =
5/3$. The slowdown in our simulation from 400 km/s at 1 AU, down to
335 km/s at the TS matches the 15 \% slowdown inferred from {\it
  Voyager} 2 observations \citep{RLandW08}.  {\it Voyager} 2 observed
a TS compression ratio of about 2 \citep{Richardson_AGU07}, which
corresponds to a Mach number of 1.7 if we assume a simple gas-dynamic
shock.  Our simulation yields a Mach number of 2.3, which is slightly
higher, due, in part, to the absence of a shock precursor. The
implications of using a $\kappa$-distribution in the heliosheath, and
how this result relates to a traditional Maxwellian approach, is
described in the next section.

\begin{table}[ht!]
\begin{center}
\begin{tabular}{|l|c|c|}
\hline
 & Maxwellian & $\kappa = 1.63$ \\
\hline
TS distance (AU)       & 83    & 87   \\
HP distance (AU)       & 139   & 131  \\
BS distance (AU)       & 400   & 440  \\
$n_H$ at TS (cm$^{-3}$) & 0.095 & 0.09 \\
$n_H$ at H-wall (cm$^{-3}$) & 0.23 & 0.215 \\
\hline
\end{tabular}
\end{center}
\caption{Comparison of global heliospheric densities and distances in
  the upstream LISM direction between the
  solution with a Maxwellian distribution for protons in the
  heliosheath, and when we take protons to obey a
  $\kappa$-distribution in the inner heliosheath with $\kappa = 1.63$
  and allow feed-back of the modified ENA distribution on the global
  solution.}
\label{table:distances}
\end{table}

\subsection{Implications of using a $\kappa$-distribution in the heliosheath}\label{sec:implications}

\begin{figure}[ht!]
\epsscale{1.0}
\plotone{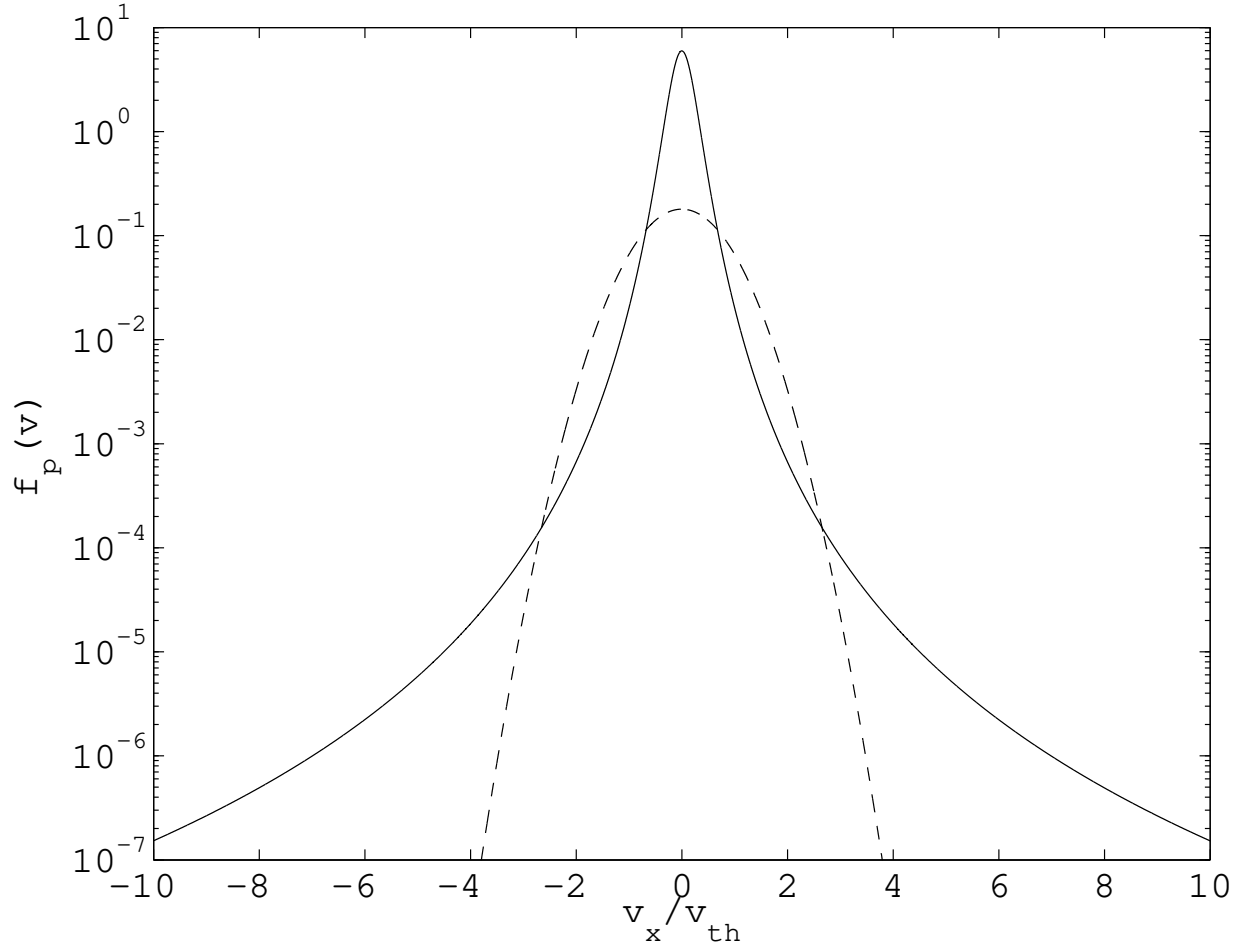}
\caption{A 1D slice of the velocity distribution function in the
  plasma frame for $\kappa = 1.63$, based on (\ref{eq:kappa_dist})
  (solid line), along with Maxwellian distribution (dashed). Note that
  the core of the $\kappa$-distribution is narrower than the
  Maxwellian. The zeroth and second moments are the same for both
  distributions. To aid comparison, we have defined $v_{th} =
  \Theta_p\sqrt{\kappa/(\kappa-3/2)}$ to the thermal speed parameter
  $\Theta_p$ of the $\kappa$-distribution, where $v_{th} = 2 k_B
  T/m_p$ is the Maxwellian thermal speed.}
\label{fig:kappa1.63_Maxwellian}
\end{figure}

Pick-up ions (PUI's) originate in the SW due to charge-exchange of
LISM neutrals with SW protons. However, they do not thermalize with
the background SW plasma \citep{Isenberg86,Zank99} and are not
therefore equilibrated with the SW.  Thus, PUI's constitute a separate
suprathermal population of the SW
\citep{MHKSandG85,GGBFGIOvSandW93,Gloeckler96,GandG98}. PUI's
contribute to the power-law tails observed almost universally in the
SW plasma distribution \citep{Mewaldt_etal01,FandG06}. A simple way to
add a power-law tail, and thereby model the proton, energetic
particle, and PUI populations as a single distribution, is to assume a
generalized Lorentzian, or ``$\kappa$'', function
\citep{BAFHandS67,SandT91,Collier95,Leubner04} given by
\begin{equation}\label{eq:kappa_dist}
f_p({\bf v}) = \frac{n_p}{\pi^{3/2}\Theta_p^3} \frac{1}{\kappa^{3/2}}
\frac{\Gamma(\kappa+1)}{\Gamma(\kappa-1/2)} \left[1 + \frac{1}{\kappa} 
\frac{({\bf v} - {\bf u}_p)^2}{\Theta_p^2}\right]^{-(\kappa+1)}
\end{equation}
where $\Theta_p$ is a typical speed related to the effective
temperature of the distribution, and is evaluated using the pressure
equation (\ref{eq:pressure}) below. This distribution has a
Maxwellian core, a power-law tail which scales as $v^{-2\kappa-2}$,
and reduces to a Maxwellian in the limit of large $\kappa$. Although
the core and tail features agree qualitatively with observations, a
limitation of the $\kappa$ formalism is that it does not allow us to
adjust their relative abundances. The observed flat-topped PUI
population is also absent in the $\kappa$ approximation. In Figure
\ref{fig:kappa1.63_Maxwellian}, we plot a $\kappa$-distribution for
$\kappa = 1.63$, along with a Maxwellian distribution.

The basic principle in our approach is to note that the MHD equations
for the plasma do not change if we assume a $\kappa$-distribution for
SW protons. This is facilitated by the fact that the basic fluid
conservation laws do not assume any specific form of the distribution
function \citep[see for example][]{Burgers69}. Closure at the second
moment is possible if the distribution is isotropic, since the heat
flux and the off-diagonal components of the stress tensor are then
identically zero. The only difference from conventional fluid dynamics
is that the collision integrals do not vanish as they would for a
Maxwellian distribution. However, collisional frequencies are so low
for the SW that we may neglect these collisional terms and treat the
distribution function (\ref{eq:kappa_dist}) as ``frozen'' into the
plasma. Even though the SW is effectively collisionless, an MHD
approach is still warranted since the plasma has fluid properties
perpendicular to the magnetic field, while various wave phenomena help
isotropize this \citep[see for example][]{Kulsrud84}. For these
reasons we solve the regular MHD equations to find the bulk plasma
quantities, but in the inner heliosheath we simply interpret these as
having come from (\ref{eq:kappa_dist}). For simplicity we assume
$\kappa = 1.63$ in all SW plasma, which is a value consistent with the
data analysis of \cite{DKRHAGHandL05}. As we show in Section
\ref{sec:ENA_spectra}, observations by the upcoming {\it IBEX} mission
can be used to estimate $\kappa$ in the heliosheath.

The two distribution functions, $\kappa$ and Maxwellian, used to model
the plasma are linked through the choice of $\Theta_p$, and we reconcile
these using the isotropic plasma pressure, given by
\begin{equation}\label{eq:pressure}
P = \frac{m_p}{3} \int_0^\infty v^2 f_p(v) \; 4\pi v^2 {\rm d}v 
= \frac{m_p n_p}{2} \Theta_p^2 \frac{\kappa}{\kappa-3/2}
\end{equation}
Note that the thermal core collapses as $\kappa \rightarrow 3/2$ and
the pressure becomes undefined. This limiting case corresponds to a
$v^{-5}$ tail \citep{FandG06}.  For the purposes of comparison, we
define an effective temperature for the $\kappa$-distribution
\begin{equation}\label{eq:T_eff}
T_{\rm eff} = \frac{P}{n_pk_B}
\end{equation}
The temperature profiles depicted in Figures
\ref{fig:HPFZandL07_global} and \ref{fig:plasma_slices} refer to the
effective temperature.

Charge-exchange couples the neutral and plasma populations. However,
the charge exchange loss terms are different when we use a
$\kappa$-distribution for protons. In the Appendix we derive the
charge exchange rate for a hydrogen atom traveling through a
$\kappa$-distribution of protons, which is used in our kinetic code
for H atoms in the heliosheath.

Other authors have included pick-up ions into their heliospheric
models in various different ways. The Bonn model \citep{FKandS00}
include PUI's as a separate fluid with a source term due to
interstellar neutrals charge-exchanging in the supersonic SW, and a
sink due to PUI's being energized and becoming part of the anomalous
cosmic ray population, which is modeled as a separate fluid. The PUI
distribution function of the Bonn model is assumed to be isotropic and
flat-topped between 0 and $v_{SW}$ in the frame of the SW. Although
this type of distribution agrees reasonably well with observations of
PUI's in the supersonic SW \citep{GandG98}, the validity of the same
distribution downstream of the TS is more questionable. Such a
distribution also does not have a tail that extends beyond the pick-up
energy, which is a requirement for obtaining ENA's at high
energies. This model was modified in \cite{FandS04} to include a
significant improvement in the form of the PUI distribution, based on
the work of \cite{FandL00} which includes analytic estimates of the
effects of upstream turbulence. Although restricted by axial symmetry,
this model includes time-dependent effects, and allows the authors to
estimate various properties of ENA's.

\cite{MIandC06} recently introduced a more complicated PUI model based
on earlier work by \cite{CFandI03}. In this model a host of different
neutral atom and PUI populations are tracked kinetically. This model
incorporates more physics than our relatively simple
$\kappa$-distribution approach, but to manage the added complexity, it
also requires a number of additional assumptions. These include the
form of the velocity diffusion coefficient, that the magnetic moment
is conserved by PUI's as they cross the TS, and an ad hoc assumption
about the downstream energy partition between electrons, protons and
PUI's. The increased computational requirements also forces
\cite{MIandC06} to consider only the case of axial symmetry, thereby
neglecting the IMF and restricting the ISMF to being aligned with the
flow. Although their assumptions are reasonable, it is difficult to
determine the influence these have on their conclusions. One of the
interesting results from their model is that the locations of the TS,
HP and BS change when the effects of PUI's are allowed to
self-consistently react back on the plasma -- a result which agrees
quite well quantitatively with our findings in the next section.

\section{Effects of heliosheath $\kappa$-distribution on the global solution}\label{sec:global_solution}

\begin{figure}[ht!]
\epsscale{1.0}
\plotone{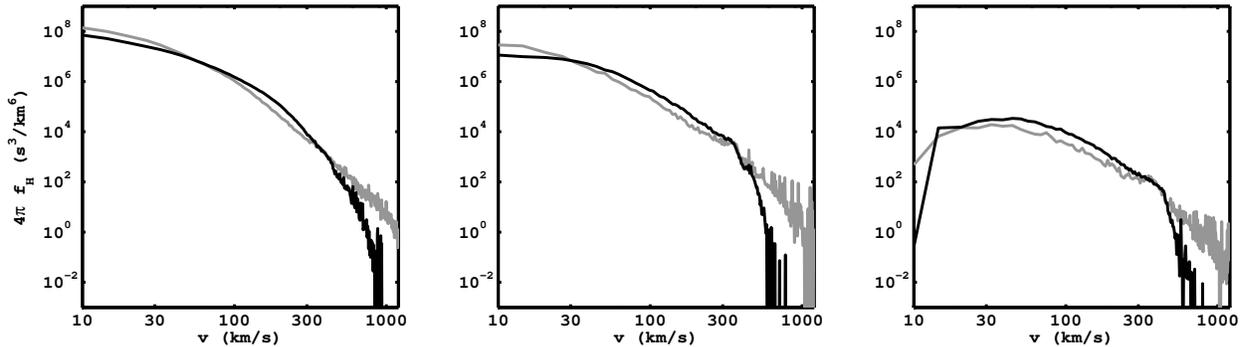}
\caption{Velocity distributions of ENA's at three locations along the
  axis defined by the LISM flow vector with the Sun at the origin:
  -400 AU in the heliotail (left), 180 AU upstream in the hydrogen
  wall (middle), 600 AU in the nearby LISM (right). The black line is
  for ENA's obtained from a Maxwellian distribution of heliosheath
  ions (the parent population of ENA's), while the gray line is
  commensurate to a $\kappa = 1.63$ distribution for heliosheath
  protons in the same steady-state configuration. Note that for small
  $\kappa$ we have less medium energy ENA's, but more at low and high
  energies, in agreement with the respective distributions shown in
  Figure \ref{fig:kappa1.63_Maxwellian}. }
\label{fig:HPFZandL07_dist}
\end{figure}

In the preceding section we showed that we may solve the regular MHD
equations for the plasma in the heliosheath, and interpret these
results in terms of a $\kappa$-distribution for the ion population. It
is less clear, however, what the effects of $\kappa$-distributed
neutral atoms originating from the heliosheath will have on the global
heliosphere-interstellar medium solution. Figure
\ref{fig:HPFZandL07_dist} shows the velocity distribution of heliosheath
hydrogen at various locations along the LISM flow vector. It is clear
from this figure that for a $\kappa = 1.63$ distribution significantly
more H-atoms with energies above 1 keV result than for a Maxwellian
ion population in the heliosheath. It is also important to note that
ENA's in the heliotail (left plot) show a clear power-law tail ($\sim
v^{-2(\kappa + 1)}$), mirroring the plasma, when a $\kappa$-distribution is
assumed for heliosheath protons. These tails persist even outside the
heliosphere (middle and right plots) for energies above 1 keV.

To test the effect of keV ENA's on the global heliosphere, we ran our
code with $\kappa = 1.63$ in the heliosheath, and allowed these ENA's
to feed back self-consistently on the global solution. Since H-atoms
are modeled kinetically, this provides no extra difficulty for our
model. The only difference, by comparison with the case of a
Maxwellian proton distribution, is that we need to use a different
formula for the relative motion between a given particle and the
ambient plasma. This formula is derived in the appendix.

Figure \ref{fig:plasma_slices} compares plasma density and temperature
along radial lines in the nose, polar and tail directions for the
Maxwellian and equilibrated $\kappa = 1.63$ heliosheath cases.
Secondary charge-exchange of neutrals created in the hot heliosheath
was identified by \cite{ZPWandH96} as a critical medium for the
anomalous transport of energy from the shocked solar wind to the
shocked and unshocked LISM. In particular, the upwind region abutting
the HP experienced considerable heating as a result of secondary
charge-exchange of hot ($\sim 10^6$ K) neutrals with the cold LISM
protons. The efficiency of this medium of anomalous heat transfer is
increased with a $\kappa$-distribution in the inner heliosheath. This
results simultaneously in a shrinking of the inner heliosheath and an
expansion of the outer heliosheath. The inner heliosheath plasma
temperature (defined in terms of pressure) remains unchanged, because
the Maxwellian and $\kappa$-distributions have the same second moment
(see Section \ref{sec:implications}). We find that in the nose
direction the termination shock moves out by about 4 AU, while the
heliopause moves inward by about 9 AU. The bow shock stand-off
distance increases by 25 AU, and the shock itself is weakened by the
additional heating of the LISM plasma by fast neutrals from the
SW. Table \ref{table:distances} summarizes these changes in heliospheric
geometry. The observed modifications to the heliospheric discontinuity locations
agree quite well with the changes observed by the multi-component
heliospheric model of \cite{MIandC06}, which includes a kinetic
representation of PUI's. These authors report a 5 AU increase in the
TS distance and a 12 AU decrease in the distance to the HP, for an
axially symmetric calculation without magnetic fields.

\begin{figure}[ht!]
\epsscale{1.0}
\plotone{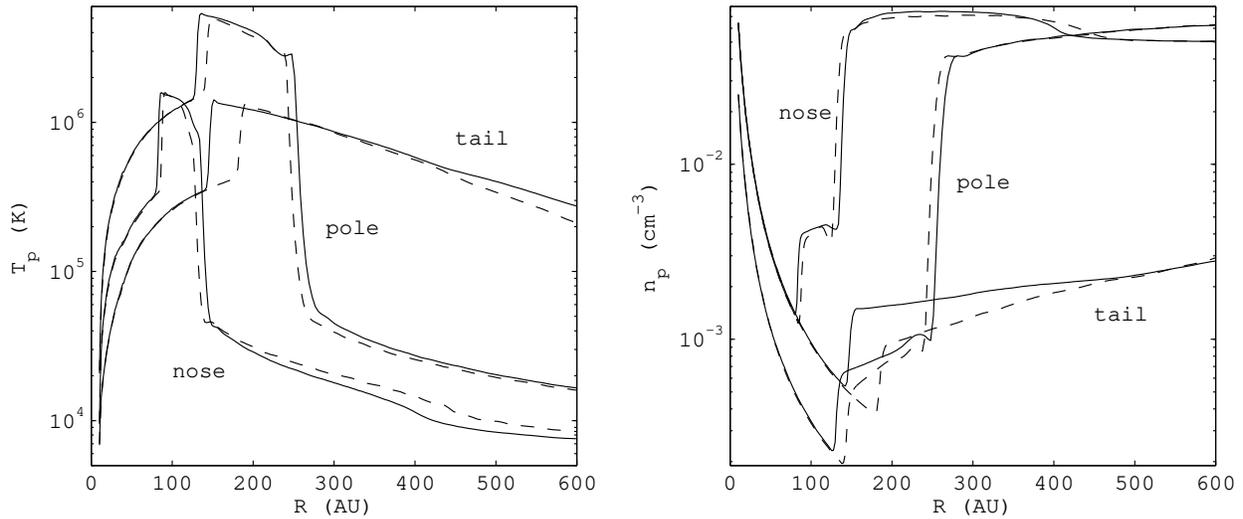}
\caption{Radial profiles of effective plasma temperature (left) and
  density (right) in the nose, polar (i.e. in the meridional plane),
  and tail directions. The solid line represents the values obtained
  by using a Maxwellian distribution function for the proton
  distribution and ENA's generated from it. The dashed line is
  obtained by assuming that the proton distribution in the supersonic
  and subsonic SW can be described as an isotropic
  $\kappa$-distribution with $\kappa = 1.63$. Although the MHD
  equations do not change in the latter case, the distribution
  function of ENA's born through charge-exchange in the heliosheath
  becomes more $\kappa$-like (see Figure \ref{fig:HPFZandL07_dist})
  and their secondary charge-exchange outside the heliosheath alters
  the global plasma configuration. The temperature plots also
    demonstrate the relationship between PUI pressure and SW speed,
    with the fast SW over the poles showing a much higher
    temperature/pressure than the slower ecliptic SW.}
\label{fig:plasma_slices}
\end{figure}

Another important distinction between the Maxwellian and
$\kappa$-distribution based models is that the filtration rate of
hydrogen changes at the heliopause. We find that in the Maxwellian
case the hydrogen density at the TS is about 63\% of the interstellar
value, while for the $\kappa$-distributed model the density drops
slightly to 60\%. As with the TS and HP locations, these results agree
quite well with the \cite{MIandC06} model.

\section{Implications for {\it IBEX}}

The Interstellar Boundary EXplorer mission will provide all-sky maps
of ENA's coming from the inner heliosheath, at 14 energy bands from 10
eV to 6 keV. However, this data is unusual in that all the ENA's
detected at a particular pixel and energy bin, will have come from a
large volume of space with non-uniform plasma properties. As such it
is not possible to invert an ENA map to determine the heliosheath's
shape, size, and plasma distribution. For this reason, we need to use
forward modeling to help us understand the relationship between model
heliosheaths and their corresponding synthetic ENA maps. In
\cite{HPZandF07a}, we identified several possible signatures to infer
heliosheath properties from {\it IBEX} data. Below we
present ENA maps and spectra from our improved heliospheric model, and
relate these to the properties of our model heliosheath.

\subsection{Ionization losses}\label{sec:ionization_losses}

ENA's propagating from the heliosheath to a detector at 1 AU may
experience re-ionization due to charge exchange, electron impact
ionization, or photo-ionization. These effects are of major importance
close to the Sun, and in the simplest approximation scale according to
\begin{equation}\label{eq:losses}
w = w_0 \exp(-\int\beta \; {\rm d}t) \;,\quad
\beta(r) = \beta_E/r^2[AU] \;,\quad
\beta_E \simeq 6\times 10^{-7} {\rm s}^{-1}
\end{equation}
where $w$ is a pseudo-particle weight which is initially equal to
$w_0$ at the point of charge-exchange and decays with time as a
function of position. Alternatively, we can view $w/w_0$ as the
survival probability for a particular particle. We note here that
$\beta_E$ does not have to be uniform in all directions, so that
ionization losses for particles coming in over the poles could be
different from those traveling in the ecliptic plane, and it may also
have temporal variations.

\begin{figure}[ht!]
\epsscale{1.0}
\plotone{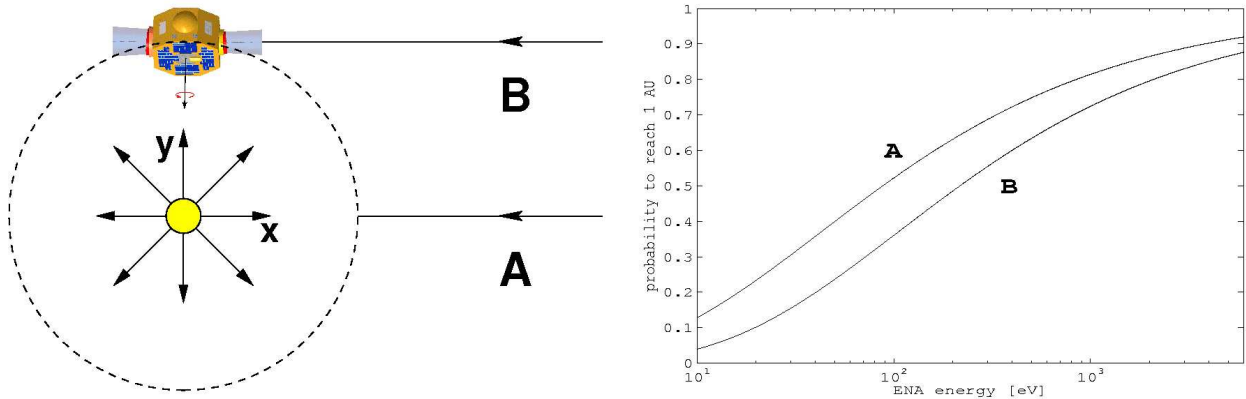}
\caption{Schematic (left) showing the difference between ENA flux at 1
  AU (dashed circle) along path A, and the ENA flux {\it IBEX} will measure
  along path B. Note that the {\it IBEX} instrument always points
  perpendicular to the radial vector from the Sun. The right plot
  shows the different survival probabilities along the two paths from
  some point in the heliosheath (effectively infinity) to 1 AU, due
  to charge-exchange, electron impact and photo-ionization losses.}
\label{fig:IBEX_path}
\end{figure}

Generally ENA's will travel on effectively straight trajectories since
solar gravity is approximately balanced by radiation pressure.
\cite{BandT06} show that for solar minimum conditions the deflection
angle will be less than 5 degrees, even for the lowest energies we
consider. In the simulations presented here, we assume zero
deflection, since we are mainly interested in the gross features of
the ENA maps. Trajectory ``A'' in Figure \ref{fig:IBEX_path} shows the
shortest straight-line path to 1 AU for an ENA, while path B
represents the longest.  If we assume straight line propagation
  at constant speed $-v_0$, then the survival probability
  (i.e. $w/w_0$) is given by
$$
P = \exp\left( -\frac{\beta_E}{v_0}\int_1^\infty
\frac{1}{x^2+y_0^2} {\rm d} x \right) \;,
$$
where $y_0 = 0$ for path A and $y_0 = 1$ for path B. Upon integration
we have
\begin{equation}
P_A = \exp\left(-\frac{\beta_E}{v_0}\right) \;,\quad
P_B = \exp\left(-\frac{\pi\beta_E}{2v_0}\right)
\end{equation}
where $v_0$ is the particle speed in AU per second. Here path B is
relevant to {\it IBEX} observations, but experiences more ionization
losses. A simple $\pi/2$ factor can be used to switch between 1 AU
fluxes and {\it IBEX} fluxes, assuming no deflection due to gravity or
radiation pressure occurs. Figure \ref{fig:IBEX_path} shows
survival probability profiles for both paths, and we note that profile
``A'' corresponds to Figure 4 of \cite{GRMFFandMcC01}. These loss
formulae will we used in the next section to undo the losses simulated
in the code so that we can use the pristine ENA fluxes to construct
energy spectra. Such a procedure would also be necessary for {\it
  IBEX} data, when we want to infer properties of the parent plasma.

\subsection{ENA spectra}\label{sec:ENA_spectra}

We may extract information about the proton energy spectrum in the
heliosheath by simply plotting the {\it IBEX} energy bin data for a
particular pixel (i.e. direction). Our global model allows us to both
prescribe a form for the distribution function in the heliosheath for
ENA's (i.e. $\kappa$) and then attempt to deconvolve this from the
data. The only difference is that {\it IBEX} spectral data will be
line-of-sight integrated, rather than at a particular point in
space. Nevertheless, we have the global data from our model, which we
can use to compare an {\it IBEX} line-of-sight spectrum with plasma
properties along that line of sight. This is particularly interesting
in the nose direction, where the plasma distribution observed by the
{\it Voyager} spacecraft can be compared with the spectral slope
inferred from the {\it IBEX} data.

\begin{figure}[ht!]
\epsscale{1.0}
\plotone{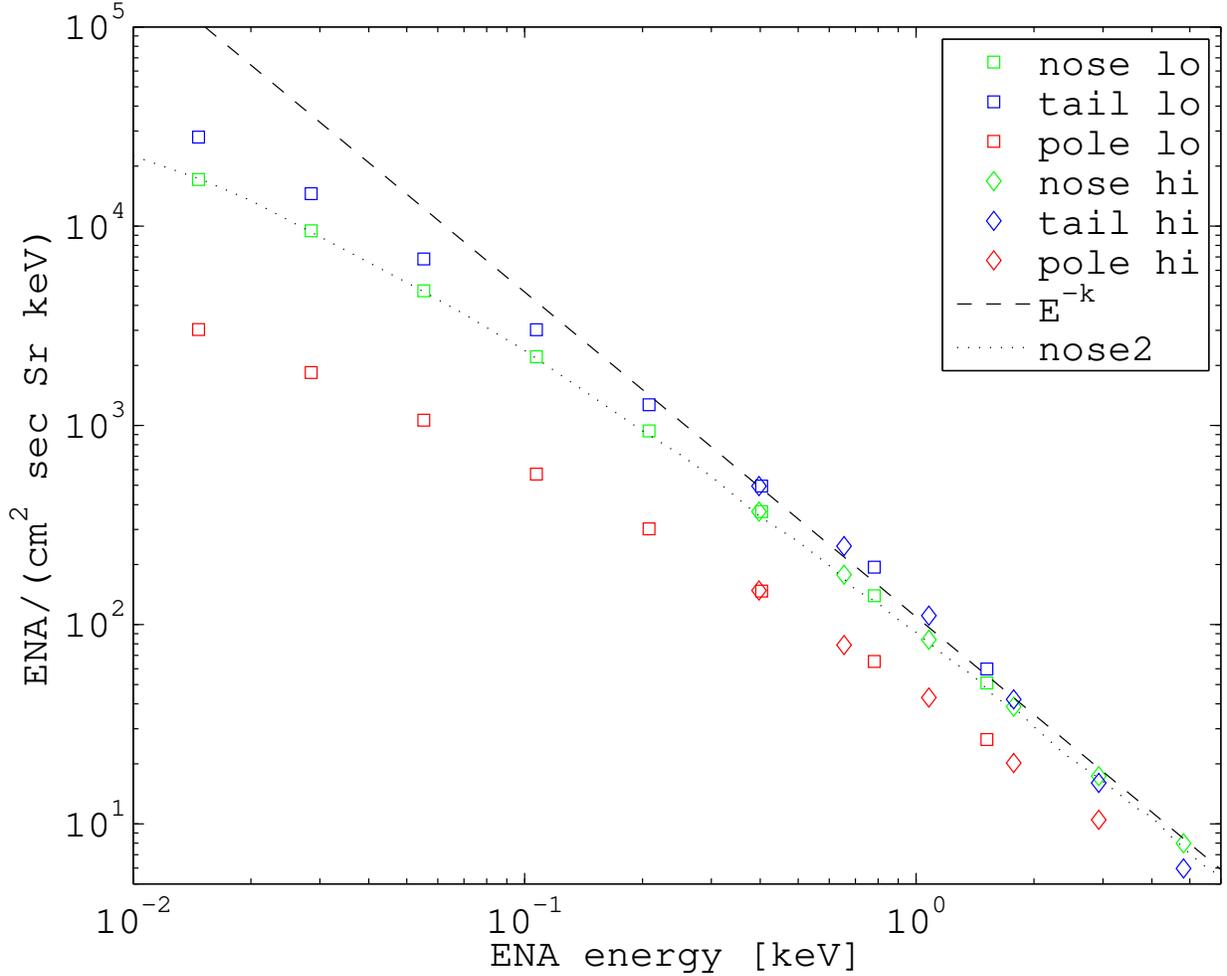}
\caption{ENA energy spectra as observed at 1 AU along various lines of
  sight. Here the squares and diamonds represent data using
  approximate {\it IBEX} energy bins obtained by dividing the {\it
    IBEX}-lo and {\it IBEX}-hi energy ranges (0.01 -- 2.0 keV and 0.3
  -- 6 keV) into 8 and 6 equal bins on a logarithmic scale
  respectively \citep[see also][]{Prested_kappa08}. The dotted line
  was obtained using narrower bins (32 total), and demonstrates that
  the {\it IBEX} bin widths are sufficiently narrow to maintain
  accuracy. The dashed line has a slope of $-\kappa$, which represents
  the plasma spectrum at a particular point, and appears reasonably
  well reproduced along the lines of sight considered.}
\label{fig:ibex_spectra}
\end{figure}

To obtain a more accurate representation of the ENA spectrum in the
heliosheath, we need to undo the ionization losses experienced by
particles as they travel to the detector. In Section
\ref{sec:ionization_losses} we derived a simple expression to estimate
the survival probability of a particle with a given energy along a
particular line of sight.  Figure \ref{fig:ibex_spectra} shows three
energy spectra for ENA's originating from the nose, tail and polar
directions. For these spectra, we have divided the flux measured at 1
AU by the survival probability for each energy band to undo the
ionization losses, as mentioned above. We find that for the three
directions considered, the energy spectrum tends toward the value of
$-\kappa$ above about 1 keV. This result shows that the {\it IBEX}
data, in spite of being line-of-sight integrated, should be able to
help determine the spectral slope of the heliosheath protons in the
0.6 -- 6 keV range.

Figure \ref{fig:ibex_spectra} also shows that the spectra in the three
directions considered have very similar properties. This will not
necessarily be true for the real heliosphere, where the post-shock SW
may develop different high energy tails in different directions. The
dotted line (labeled ``nose2'') is for a spectrum in the nose
direction obtained using 32 energy bins (compared to about 10
non-overlapping {\it IBEX} bins). The agreement between this curve and
the green markers shows that, for $\kappa = 1.63$ at least, the number
of {\it IBEX} bins is sufficient to reproduce the spectrum.

\subsection{ENA all-sky maps}

The method we use for computing all-sky ENA maps is described in
\cite{HPZandF07a}, where we first obtain a steady-state heliosphere
and then trace ENA's born through charge-exchange in the heliosheath
down to 1 AU, where these are then binned according to energy and the
direction of origin. Additional ionization losses along the particle's
trajectory act to ``evaporate'' its computational weight. The key
difference from our previous results is that we now assume a
$\kappa$-distribution for the heliosheath protons which form the
parent population for ENA's. This modification allows us to obtain
ENA's up to several keV, and is more consistent with SW data.

\begin{figure}[ht!]
\epsscale{1.0}
\plotone{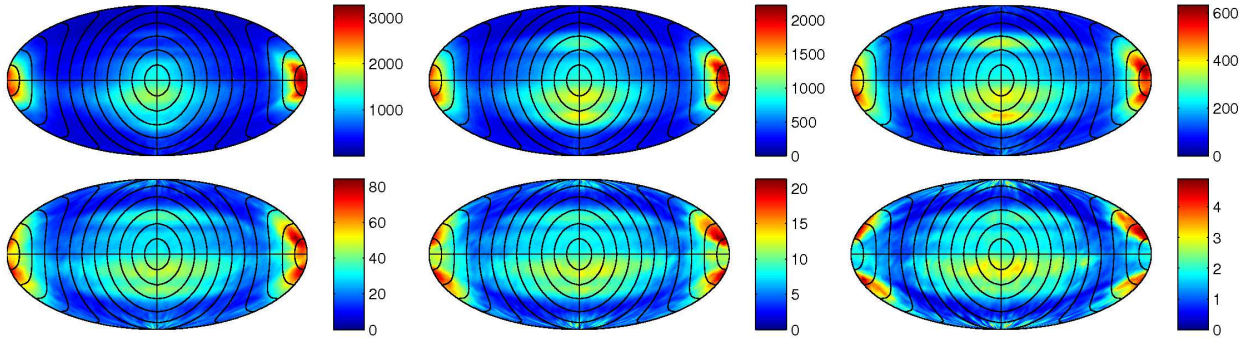}
\caption{All-sky maps of energetic neutral atom flux at 1 AU, in units
  of (cm$^2$ sr s keV)$^{-1}$, generated in the inner heliosheath
  through charge-exchange between an interstellar neutral atom and a
  heliosheath proton drawn from a $\kappa$-distribution with $\kappa =
  1.63$. The direction of the LISM flow is at the center of
  the plot, with the poles top and bottom, and the heliotail on the
  far sides. Contour lines have been drawn at 15 degrees
  intervals. Maps are generated by binning ENA's which intersect the 1
  AU sphere on radially inward trajectories. The maps shown are for
  the following energies and bin-widths (in eV): $10 \pm 2$, $50 \pm
  10$, $200 \pm 20$, $1000 \pm 100$, $2400 \pm 200$, and $6000 \pm
  400$ (from top left to bottom right). }
\label{fig:skymaps}
\end{figure}

Figure \ref{fig:skymaps} shows all-sky ENA maps obtained from our
steady-state solution with a $\kappa$-distribution for heliosheath
protons. The top right plot shows the ENA map for 200 eV, which can be
compared with our previous work \citep{HPZandF07a}, where we did not
self-consistently couple the plasma and kinetic neutral atoms, and
where we assumed a Maxwellian proton distribution. We find that when
we use a $\kappa$-distribution, the ENA flux at 200 eV is two to three
times smaller than for the Maxwellian case, due to the shape of the
proton distribution (see Figure \ref{fig:kappa1.63_Maxwellian}) and
resulting ENA distribution (Figure \ref{fig:HPFZandL07_dist}), as well
as the thinner inner heliosheath resulting from the use of a
$\kappa$-distribution (see Section \ref{sec:global_solution}). As
expected, this decrease of medium energy (100's of eV) ENA's is
compensated by an increased ENA flux above 1 keV. Our results predict
a count rate of about 3 atoms per (cm$^2$ sr s keV) at 6 keV.

Less obvious is the decline in low energy flux when compared to the
Maxwellian results \citep{HPZandF07a}, even though there are more
ENA's being generated at the lowest energies (see Figure
\ref{fig:HPFZandL07_dist}). The principal reason for this is that the
SW core temperature is significantly lower when we use $\kappa$, so
that these ENA's lack the energy to propagate upstream, since the bulk
speed exceeds the thermal speed of the core. This low SW core
temperature is in fact qualitatively consistent with the latest {\it
  Voyager} 2 findings \citep{Richardson_AGU07}.

The heliosphere depicted in Figure \ref{fig:HPFZandL07_global}, is
commensurate to approximately ``solar minimum'' conditions, with a
clearly defined high speed wind emanating from the poles. The high
speed wind gives rise to hotter high latitude heliosheath plasma,
which in turn increases the energy of ENA's generated in the subsonic
polar SW. The all-sky maps of Figure \ref{fig:skymaps} show that at
energies above about 1 keV, these streams of hot SW dominate the ENA
flux, while at lower energies the central tail region is the major
source of ENA's.

Comparing skymaps at different energies, we see from Figure
\ref{fig:skymaps} that the qualitative properties do not vary widely
over the {\it IBEX} energy range. This contrasts sharply with the
results for a Maxwellian heliosheath, where we generally see a higher
flux coming from the tail than the nose at low energies, and the
reverse at high energies \citep{HPZandF07a}. This can be attributed to
the steep decline in the Maxwellian distribution, compared to the much
broader $\kappa$-distribution (see Figure
\ref{fig:kappa1.63_Maxwellian}), which means that particles observed
at a given energy have come from plasma with a narrower range of
temperatures. In other words, the relatively cool plasma in the
distant heliotail can still be a significant source of high energy
ENA's, if we assume it has a $\kappa$-distribution. Only at the
highest energies, above about 2 keV, does the nose-tail asymmetry
favor the nose direction.

\section{Conclusions}

We have used our 3D MHD-kinetic code to investigate the impact of
assuming an alternative heliosheath proton distribution, a
$\kappa$-distribution rather than the more usual Maxwellian, on both
the SW-LISM interaction region, and the observed ENA flux at 1 AU. The
motivation for this is that pick-up ions, generated when an
interstellar neutral atom charge-exchanges in the supersonic solar
wind, form high energy tails that are always observed in the solar
wind plasma. The $\kappa$-distribution has core and tail features, and
is often invoked in data analysis of the SW proton distribution
function. The use of a $\kappa$-distribution introduces (possibly)
more realistic estimates of the ENA flux at 1 AU, and thereby serves
as an important tool in reconciling global heliospheric models with
data from the upcoming {\it IBEX} mission. One drawback of this
approach is that we cannot control the ratio between core and tail
populations. While obviously not capturing the full details of the
thermal and PUI plasma distributions in either the inner heliosheath
or throughout the supersonic SW, a $\kappa$-distribution is
nonetheless well grounded in observations as a general representation
of the SW distribution function.

We used $\kappa=1.63$ in our calculations, based on the {\it Voyager}
1 LECP data of \cite{DKRHAGHandL05}. Although the LECP data is for
much higher energies than {\it IBEX} will measure, we have shown that
{\it IBEX} data can be used to infer the spectral slope of the
heliosheath distribution for energies between 1 keV and 6 keV. The
tails of the energy spectra may have different slopes in different
directions (over the poles, for example).

The use of a $\kappa$-distribution for the ENA parent proton
population results in a significant increase of the ENA flux at
energies above 1 keV, when compared with a Maxwellian
distribution. Our results predict a count rate of about 3 per (cm$^2$
sr s keV) at the highest energies considered by {\it IBEX}, which is
many orders of magnitude higher than could be expected from a
Maxwellian heliosheath distribution. At the same time, there is a
marked reduction in the flux for intermediate energies, to about half
the Maxwellian value at a few hundred eV.  We have also calculated the
feed back of the revised ENA distribution on the global heliospheric
solution. The result is an increased transport of energy from the
inner to the outer heliosheath, with a corresponding thinning and
expansion of the former and latter. The distance between the TS and HP
decreases by 13 AU (about 25\%) in the nose direction, and the bow
shock moves out farther and becomes very weak. The thinner heliosheath
is also partly responsible for the decreased ENA flux at energies of a
few hundred eV.

Finally, we note that we have not considered time-dependent effects in
this paper. \cite{SFandS07} recently looked at the changes in the ENA
maps when they included a simple model for the solar cycle into their
3D hydrodynamic (i.e. no magnetic fields) code which includes a single
fluid for neutral gas. They found cyclic changes in the ENA flux at
100 eV, which varied by about 25\%. The observed variations at 1 keV
were considerably larger, but because they assumed a Maxwellian
distribution for protons in the heliosheath, their fluxes were about
an order of magnitude lower than ours at this energy. Effectively,
they found that fluctuations in ENA flux due to the solar cycle are
relatively small for energies close to the core of the distribution (a
few hundred eV in the heliosheath), while at high energies the changes
in ENA flux are larger. Since the $\kappa$-distribution declines much
more slowly than the Maxwellian away from the core, we expect our ENA
fluxes to vary by perhaps 50\% over a solar cycle for energies
relevant to {\it IBEX}. This, however, remains to be confirmed.

\acknowledgements 
This work was supported by NASA grants NNG05GD45G,
NNG06GD48G, and NNG06GD43G, and NSF award ATM-0296114.  Calculations
were performed on supercomputers Fujitsu Primepower HPC2500, in the
framework of the collaborative agreement with the Solar-Terrestrial
Environment Laboratory of Nagoya University, Columbia at NASA Ames
Research Center (award SMD-06-0167), and IBM Data Star (award
ATM-070011) in the San Diego Supercomputer Center.

\section*{{\bf Appendix:} Charge-exchange formulation with a $\kappa$- distribution}

Our kinetic neutral atom method solves the time-dependent Boltzmann equation
\begin{equation}\label{eq:Boltzmann}
\frac{\partial}{\partial t}f_H + {\bf v}\cdot\nabla f_H +
\frac{\bf F}{m_p}\nabla_{\bf v}\cdot f_H = P - L \;,
\end{equation}
using a Monte Carlo approach. Here $f_H$ is the distribution function
of neutral hydrogen, ${\bf F}$ is the external force, and $P$ and $L$
are the production and loss terms. Below we derive the loss rate for a
neutral particle traveling through a $\kappa$-distribution of protons.

The production and loss rates for the hydrogen population may be
written as
\begin{equation}
P = f_p({\bf x},{\bf v},t)\eta({\bf x},{\bf v},t) \;,
\end{equation}
\begin{equation}
L = f_H({\bf x},{\bf v},t)\beta({\bf x},{\bf v},t) \;,
\end{equation}
where
\begin{equation}\label{eq:eta}
\eta({\bf x},{\bf v},t) = \int \sigma_{ex} f_H({\bf x},{\bf v}_H,t)
\;|{\bf v} - {\bf v}_H| \;{\rm d}{\bf v}_H
\end{equation}
\begin{equation}\label{eq:beta1}
\beta({\bf x},{\bf v},t) = \int \sigma_{ex} f_p({\bf x},{\bf v}_p,t)
\;|{\bf v} - {\bf v}_p| \;{\rm d}{\bf v}_p \;.
\end{equation}
Here we assume that the charge exchange cross-section, approximated using
the \cite{FSandS62} expression 
\begin{equation}\label{eq:sigma_ex}
\sigma_{ex}(v_{rel}) = \left[2.1 - 0.092\ln(v_{rel})\right]^2 10^{-14}
      {\rm cm}^2 \;,
\end{equation}
varies slowly and can be taken outside the integrals in (\ref{eq:eta})
and(\ref{eq:beta1}). 

In the kinetic code we require the neutral loss term $\beta$ to
compute charge-exchange on a particle-by-particle basis. To derive
this, we use the $\kappa$-distribution for the charged component, i.e.,
\begin{equation}
f_p({\bf v}_p)=\frac{n_p}{\pi^{3/2}\Theta_p^3}\frac{1}{\kappa^{3/2}}
\frac{\Gamma(\kappa+1)}{\Gamma(\kappa-1/2)}\left[1+\frac{1}{\kappa}
\frac{({\bf v}_p-{\bf u}_p)^2}{\Theta_p^2}\right]^{-(\kappa+1)},
\end{equation}
where ${\bf u}_p$ is the bulk speed and $\Theta_p$ is related to the
plasma pressure via equation (\ref{eq:pressure}).

Upon introduction of the new variables ${\bf g}=({\bf v}-{\bf v}_p)
/(\sqrt{\kappa}\Theta_p)$ and ${\bf x}=({\bf u}_p-{\bf v}_p)
/(\sqrt{\kappa}\Theta_p)$, equation (\ref{eq:beta1}) becomes
\begin{eqnarray}
\beta=\frac{n_p \sigma_{ex} \Theta_p}{\pi^{3/2}}
\frac{\sqrt{\kappa}\Gamma(\kappa+1)}{\Gamma(\kappa-1/2)}
\int g[1+({\bf g}-{\bf x})^2]^{-(\kappa+1)}d^3g \nonumber \\
=\frac{2n_p \sigma_{ex}\Theta_p}{\sqrt{\pi}}\frac{\sqrt{\kappa}\Gamma(\kappa+1)}
{\Gamma(\kappa-1/2)}\int_0^{\infty}dg\int_{-1}^1 d\mu\;g^3
(1+g^2-2\mu g x+x^2)^{-(\kappa+1)},
\end{eqnarray}
where $\mu=\cos\theta$, $\theta$ being the angle between ${\bf g}$ and
${\bf x}$.
After integrating over $\mu$ the result is
\begin{equation}
\beta=\frac{n_p \sigma_{ex}\Theta_p}{\sqrt{\pi\kappa}x}
\frac{\Gamma(\kappa+1)}{\Gamma(\kappa-1/2)}\int_0^{\infty}
g^2\left\lbrace[1+(g-x)^2]^{-\kappa}-[1+(g+x)^2]^{-\kappa}\right\rbrace dg.
\end{equation}
Introducing the new variable $z=g-x$ in the first term and $z=g+x$ in the second
term and using the symmetry properties of the integrand, we obtain
\begin{eqnarray}\label{eq:beta2}
\beta=\frac{2n_p \sigma_{ex}\Theta_p}{\sqrt{\pi\kappa}x}
\frac{\Gamma(\kappa+1)}{\Gamma(\kappa-1/2)}
\left(\int_0^x z^2(1+z^2)^{-\kappa}dz
+x^2\int_0^x(1+z^2)^{-\kappa}dz
\right. \nonumber \\
\left.
+2x\int_x^{\infty}z(1+z^2)^{-\kappa}dz\right).
\end{eqnarray}
The integrals are
\begin{equation}\label{eq:int1}
x^2\int_0^x(1+z^2)^{-\kappa}dz
=x^3\,{_2F}_1\left(\frac{1}{2},\kappa;\frac{3}{2};-x^2\right)
=x^3(1+x^2)^{-\kappa}\,{_2F}_1\left(1,\kappa;\frac{3}{2};
\frac{x^2}{1+x^2}\right),
\end{equation}
\begin{equation}\label{eq:int2}
2x\int_x^{\infty}z(1+z^2)^{-\kappa}dz=\frac{x(1+x^2)^{-k+1}}{(k-1)},
\end{equation}
\begin{equation}\label{eq:int3}
\int_0^x z^2(1+z^2)^{-\kappa}dz=\frac{x^3}{3}\,
{_2F}_1\left(\frac{3}{2},\kappa;\frac{5}{2};-x^2\right)
=\frac{x^3}{3}(1+x^2)^{-\kappa}\,{_2F}_1\left(1,\kappa;\frac{5}{2};
\frac{x^2}{1+x^2}\right),
\end{equation}
where $_2F_1$ is the hypergeometric function.
The exact solution for $\beta$ is therefore
\begin{eqnarray}\label{eq:beta3}
\beta=\frac{2n_p \sigma_{ex}\Theta_p}{\sqrt{\pi\kappa}}
\frac{\Gamma(\kappa+1)}{\Gamma(\kappa-1/2)}(1+x^2)^{-\kappa}
\left[x^2\,{_2F}_1\left(1,\kappa;\frac{3}{2};\frac{x^2}{1+x^2}\right)\right.
\nonumber \\
\left.+\frac{x^2}{3}\,{_2F}_1\left(1,\kappa;\frac{5}{2};\frac{x^2}{1+x^2}\right)
+\frac{1+x^2}{\kappa-1}\right].
\end{eqnarray}
However, it is more convenient to take the limits $\sqrt{\kappa}x\ll
1$ and $\sqrt{\kappa}x\gg 1$ in (\ref{eq:int1}) and (\ref{eq:int3})
before the integration.  In the former limit we obtain
\begin{equation}
x^2\int_0^x(1+z^2)^{-\kappa}dz\simeq x^3,
\end{equation}
\begin{equation}
\int_0^x z^2(1+z^2)^{-\kappa}dz\simeq\frac{x^3}{3}
\end{equation}
and the expression inside the parentheses in (\ref{eq:beta2}) becomes
$x/(\kappa-1)+x^3/3$.  Finally, in this limit
\begin{equation}
\beta=\frac{2n_p \sigma_{ex}\Theta_p}{\sqrt{\pi\kappa}}
\frac{\Gamma(\kappa+1)}{\Gamma(\kappa-1/2)}\left[\frac{1}{\kappa-1}
+\frac{({\bf v}_p-{\bf u}_p)^2}{3\kappa \Theta_p^2}\right].
\end{equation}
For large $\kappa$, $\Gamma(\kappa+a)\simeq\kappa^a\Gamma(\kappa)$ and
\begin{equation}
\beta\simeq\frac{2n_p \sigma_{ex}\Theta_p}{\sqrt{\pi}}
\left[1+\frac{({\bf v}-{\bf u}_p)^2}{3\Theta_p^2}\right].
\end{equation}
In the limit $x\gg 1$ we obtain
\begin{equation}
x^2\int_0^{\infty}(1+z^2)^{-\kappa}dz=\frac{\sqrt{\pi}\Gamma(\kappa-1/2)x^2}
{2\Gamma(\kappa)},
\end{equation}
\begin{equation}
\int_0^{\infty}z^2(1+z^2)^{-\kappa}dz=\frac{\sqrt{\pi}\Gamma(\kappa-3/2)}
{4\Gamma(\kappa)}.
\end{equation}
In this limit
\begin{equation}
\beta\simeq n_p\sigma_{ex}|{\bf v}-{\bf u}_p|
\end{equation}
and is independent of $\kappa$.  A reasonable approximation to
(\ref{eq:beta3}) that has the correct asymptotic behavior is
\begin{equation}
\beta\simeq n_p\sigma_{ex}\sqrt{\frac{4\Gamma^2(\kappa+1)\Theta_p^2}
{\pi\kappa(\kappa-1)^2\Gamma^2(\kappa-1/2)}+({\bf v}-{\bf u}_p)^2}.
\end{equation}

For large $\kappa$ this reduces to the Maxwellian limit obtained by
\cite{PZandW95}
\begin{equation}
\beta\simeq n_p\sigma_{ex}\sqrt{\frac{4}{\pi}\Theta_{p}^2+({\bf v}-{\bf u}_p)^2}.
\end{equation}


\begin{thebibliography}{}

\bibitem[\protect\citeauthoryear{Alexashov \& Izmodenov}{Alexashov \&
  Izmodenov}{2005}]{AandI05}
Alexashov, D.,  \& Izmodenov, V. 2005, Astron. Astrophys., 439, 1171

\bibitem[\protect\citeauthoryear{{Bame} et~al.}{{Bame}
  et~al.}{1967}]{BAFHandS67}
{Bame}, S.~J., {Asbridge}, J.~R., {Felthauser}, H.~E., {Hones}, E.~W.,  \&
  {Strong}, I.~B. 1967, \jgr, 72, 113

\bibitem[\protect\citeauthoryear{Baranov \& Malama}{Baranov \&
  Malama}{1993}]{BandM93}
Baranov, V.B.,  \& Malama, Yu.~G. 1993, J. Geophys. Res., 98, 15157

\bibitem[\protect\citeauthoryear{{Burgers}}{{Burgers}}{1969}]{Burgers69}
{Burgers}, J.~M. 1969, {Flow Equations for Composite Gases} (Flow Equations for
  Composite Gases, New York: Academic Press, 1969)

\bibitem[\protect\citeauthoryear{{Burlaga} et~al.}{{Burlaga}
  et~al.}{2005}]{BNALCSandMcD05}
{Burlaga}, L.~F., {Ness}, N.~F., {Acu{\~n}a}, M.~H., {Lepping}, R.~P.,
  {Connerney}, J.~E.~P., {Stone}, E.~C.,  \& {McDonald}, F.~B. 2005, Science,
  309, 2027

\bibitem[\protect\citeauthoryear{{Bzowski} \& {Tarnopolski}}{{Bzowski} \&
  {Tarnopolski}}{2006}]{BandT06}
{Bzowski}, M.,  \& {Tarnopolski}, S. 2006, in American Institute of Physics
  Conference Series, Vol. 858, Physics of the Inner Heliosheath, ed.
  J.~{Heerikhuisen}, V.~{Florinski}, G.~P. {Zank}, \& N.~V. {Pogorelov}, 251

\bibitem[\protect\citeauthoryear{{Chalov}, {Fahr}, \& {Izmodenov}}{{Chalov}
  et~al.}{2003}]{CFandI03}
{Chalov}, S.~V., {Fahr}, H.~J.,  \& {Izmodenov}, V.~V. 2003, \jgr, 108, 1266

\bibitem[\protect\citeauthoryear{{Chashei}, {Fahr}, \& {Lay}}{{Chashei}
  et~al.}{2003}]{CFandL03}
{Chashei}, I.~V., {Fahr}, H.~J.,  \& {Lay}, G. 2003, Annales Geophysicae, 21,
  1405

\bibitem[\protect\citeauthoryear{{Collier}}{{Collier}}{1995}]{Collier95}
{Collier}, M.~R. 1995, \grl, 22, 2673

\bibitem[\protect\citeauthoryear{{Decker} et~al.}{{Decker}
  et~al.}{2005}]{DKRHAGHandL05}
{Decker}, R.~B., {Krimigis}, S.~M., {Roelof}, E.~C., {Hill}, M.~E.,
  {Armstrong}, T.~P., {Gloeckler}, G., {Hamilton}, D.~C.,  \& {Lanzerotti},
  L.~J. 2005, Science, 309, 2020

\bibitem[\protect\citeauthoryear{Fahr, Kausch, \& Scherer}{Fahr
  et~al.}{2000}]{FKandS00}
Fahr, H.~J., Kausch, T.,  \& Scherer, H. 2000, Astron. Astrophys., 357, 268

\bibitem[\protect\citeauthoryear{{Fahr} \& {Lay}}{{Fahr} \&
  {Lay}}{2000}]{FandL00}
{Fahr}, H.~J.,  \& {Lay}, G. 2000, Astron. Astrophys., 356, 327

\bibitem[\protect\citeauthoryear{{Fahr} \& {Scherer}}{{Fahr} \&
  {Scherer}}{2004}]{FandS04}
{Fahr}, H.-J.,  \& {Scherer}, K. 2004, Astrophys. Space Sci. Trans., 1, 3

\bibitem[\protect\citeauthoryear{{Fisk} \& {Gloeckler}}{{Fisk} \&
  {Gloeckler}}{2006}]{FandG06}
{Fisk}, L.~A.,  \& {Gloeckler}, G. 2006, \apjl, 640, L79

\bibitem[\protect\citeauthoryear{Fite, Smith, \& Stebbings}{Fite
  et~al.}{1962}]{FSandS62}
Fite, W.L., Smith, A. C.~H.,  \& Stebbings, R.~F. 1962, Proc. R. Soc. London
  Ser. A, 268, 527

\bibitem[\protect\citeauthoryear{{Gloeckler}}{{Gloeckler}}{1996}]{Gloeckler96}
{Gloeckler}, G. 1996, Space Science Reviews, 78, 335

\bibitem[\protect\citeauthoryear{{Gloeckler}, {Fisk}, \&
  {Lanzerotti}}{{Gloeckler} et~al.}{2005}]{GFandL05}
{Gloeckler}, G., {Fisk}, L.~A.,  \& {Lanzerotti}, L.~J. 2005, in ESA Special
  Publication, Vol. 592, ESA Special Publication

\bibitem[\protect\citeauthoryear{{Gloeckler} \& {Geiss}}{{Gloeckler} \&
  {Geiss}}{1998}]{GandG98}
{Gloeckler}, G.,  \& {Geiss}, J. 1998, Space Science Reviews, 86, 127

\bibitem[\protect\citeauthoryear{{Gloeckler} et~al.}{{Gloeckler}
  et~al.}{1993}]{GGBFGIOvSandW93}
{Gloeckler}, G., et~al. 1993, Science, 261, 70

\bibitem[\protect\citeauthoryear{{Gosling} et~al.}{{Gosling}
  et~al.}{1981}]{GABFZPSandH81}
{Gosling}, J.~T., {Asbridge}, J.~R., {Bame}, S.~J., {Feldman}, W.~C., {Zwickl},
  R.~D., {Paschmann}, G., {Sckopke}, N.,  \& {Hynds}, R.~J. 1981, \jgr, 86, 547

\bibitem[\protect\citeauthoryear{Gruntman et~al.}{Gruntman
  et~al.}{2001}]{GRMFFandMcC01}
Gruntman, M., Roelof, E.~C., Mitchell, D.~G., Fahr, H.~J., Funsten, H.~O.,  \&
  McComas, D.~J. 2001, J. Geophys. Res., 106, 15767

\bibitem[\protect\citeauthoryear{Heerikhuisen, Florinski, \& Zank}{Heerikhuisen
  et~al.}{2006}]{HFandZ06}
Heerikhuisen, J., Florinski, V.,  \& Zank, G.~P. 2006, J. Geophys. Res., 111,
  A06110

\bibitem[\protect\citeauthoryear{Heerikhuisen et~al.}{Heerikhuisen
  et~al.}{2006}]{igpp_conf6}
Heerikhuisen, J., Florinski, V., Zank, G.~P.,  \& Pogorelov, N.~V.(editors).
  2006, Physics of the Inner Heliosheath (AIP)

\bibitem[\protect\citeauthoryear{{Heerikhuisen} et~al.}{{Heerikhuisen}
  et~al.}{2007}]{HPZandF07a}
{Heerikhuisen}, J., {Pogorelov}, N.~V., {Zank}, G.~P.,  \& {Florinski}, V.
  2007, \apjl, 655, L53

\bibitem[\protect\citeauthoryear{{Holzer}}{{Holzer}}{1972}]{Holzer72}
{Holzer}, T.~E. 1972, J. Geophys. Res., 77, 5407

\bibitem[\protect\citeauthoryear{{Isenberg}}{{Isenberg}}{1986}]{Isenberg86}
{Isenberg}, P.~A. 1986, \jgr, 91, 9965

\bibitem[\protect\citeauthoryear{{Khabibrakhmanov} et~al.}{{Khabibrakhmanov}
  et~al.}{1996}]{KSZandP96}
{Khabibrakhmanov}, I.~K., {Summers}, D., {Zank}, G.~P.,  \& {Pauls}, H.~L.
  1996, \apj, 469, 921

\bibitem[\protect\citeauthoryear{{Kulsrud}}{{Kulsrud}}{1984}]{Kulsrud84}
{Kulsrud}, R.~M. 1984, in Basic Plasma Physics: Selected Chapters, Handbook of
  Plasma Physics, Volume 1, ed. A.~A. {Galeev} \& R.~N. {Sudan}, 115

\bibitem[\protect\citeauthoryear{{Lee} \& {Ip}}{{Lee} \& {Ip}}{1987}]{LandI87}
{Lee}, M.~A.,  \& {Ip}, W.-H. 1987, \jgr, 92, 11041

\bibitem[\protect\citeauthoryear{{Leubner}}{{Leubner}}{2004}]{Leubner04}
{Leubner}, M.~P. 2004, Physics of Plasmas, 11, 1308

\bibitem[\protect\citeauthoryear{{Malama}, {Izmodenov}, \& {Chalov}}{{Malama}
  et~al.}{2006}]{MIandC06}
{Malama}, Y.~G., {Izmodenov}, V.~V.,  \& {Chalov}, S.~V. 2006, \aap, 445, 693

\bibitem[\protect\citeauthoryear{{Matthaeus} et~al.}{{Matthaeus}
  et~al.}{1999}]{MZSandO99}
{Matthaeus}, W.~H., {Zank}, G.~P., {Smith}, C.~W.,  \& {Oughton}, S. 1999,
  Physical Review Letters, 82, 3444

\bibitem[\protect\citeauthoryear{{McComas} et~al.}{{McComas}
  et~al.}{2006}]{McComas_IGPP06}
{McComas}, D., et~al. 2006, in Physics of the Inner Heliosheath, ed.
  J.~Heerikhuisen, V.~Florinski, G.~P. Zank, \& N.~V. Pogorelov, Vol. 858
  (AIP), 400

\bibitem[\protect\citeauthoryear{{McComas} et~al.}{{McComas}
  et~al.}{2004}]{McComas_IGPP04}
{McComas}, D., et~al. 2004, in Physics of the Outer Heliosphere, ed.
  V.~Florinski, N.~V. Pogorelov, \& G.~P. Zank, Vol. 719 (AIP), 162

\bibitem[\protect\citeauthoryear{{McComas} et~al.}{{McComas}
  et~al.}{2000}]{McComas_etal00_Ulysses}
{McComas}, D.~J., et~al. 2000, \jgr, 105, 10419

\bibitem[\protect\citeauthoryear{{Mewaldt} et~al.}{{Mewaldt}
  et~al.}{2001}]{Mewaldt_etal01}
{Mewaldt}, R.~A., et~al. 2001, in American Institute of Physics Conference
  Series, Vol. 598, Joint SOHO/ACE workshop ''Solar and Galactic Composition'',
  ed. R.~F. {Wimmer-Schweingruber}, 165

\bibitem[\protect\citeauthoryear{{Moebius} et~al.}{{Moebius}
  et~al.}{1985}]{MHKSandG85}
{Moebius}, E., {Hovestadt}, D., {Klecker}, B., {Scholer}, M.,  \& {Gloeckler},
  G. 1985, \nat, 318, 426

\bibitem[\protect\citeauthoryear{Opher, Stone, \& Liewer}{Opher
  et~al.}{2006}]{OSandL06}
Opher, M., Stone, E.~C.,  \& Liewer, P.~C. 2006, Astrophys. J., 640, L71

\bibitem[\protect\citeauthoryear{{Pauls} \& {Zank}}{{Pauls} \&
  {Zank}}{1996}]{PandZ96}
{Pauls}, H.~L.,  \& {Zank}, G.~P. 1996, \jgr, 101, 17081

\bibitem[\protect\citeauthoryear{{Pauls} \& {Zank}}{{Pauls} \&
  {Zank}}{1997}]{PandZ97}
{Pauls}, H.~L.,  \& {Zank}, G.~P. 1997, \jgr, 102, 19779

\bibitem[\protect\citeauthoryear{Pauls, Zank, \& Williams}{Pauls
  et~al.}{1995}]{PZandW95}
Pauls, H.~L., Zank, G.~P.,  \& Williams, L.~L. 1995, J. Geophys. Res., 100,
  21,595

\bibitem[\protect\citeauthoryear{{Pogorelov}, {Heerikhuisen}, \&
  {Zank}}{{Pogorelov} et~al.}{2008}]{PHandZ08}
{Pogorelov}, N.~V., {Heerikhuisen}, J.,  \& {Zank}. 2008, \apjl, 675, in press

\bibitem[\protect\citeauthoryear{{Pogorelov} et~al.}{{Pogorelov}
  et~al.}{2007}]{PSFandZ07}
{Pogorelov}, N.~V., {Stone}, E.~C., {Florinski}, V.,  \& {Zank}, G.~P. 2007,
  \apj, 668, 611

\bibitem[\protect\citeauthoryear{{Pogorelov} \& {Zank}}{{Pogorelov} \&
  {Zank}}{2006}]{PandZ06b}
{Pogorelov}, N.~V.,  \& {Zank}, G.~P. 2006, in Astronomical Society of the
  Pacific Conference Series, Vol. 359, Numerical Modeling of Space Plasma
  Flows, ed. G.~P. {Zank} \& N.~V. {Pogorelov}, 184

\bibitem[\protect\citeauthoryear{Pogorelov, Zank, \& Ogino}{Pogorelov
  et~al.}{2004}]{PZandO04}
Pogorelov, N.~V., Zank, G.~P.,  \& Ogino, T. 2004, Astrophys. J., 614, 1007

\bibitem[\protect\citeauthoryear{Pogorelov, Zank, \& Ogino}{Pogorelov
  et~al.}{2006}]{PZandO06}
Pogorelov, N.~V., Zank, G.~P.,  \& Ogino, T. 2006, Astrophys. J., 644, 1299

\bibitem[\protect\citeauthoryear{Prested et~al.}{Prested
  et~al.}{2008}]{Prested_kappa08}
Prested, C., et~al. 2008, \jgr, accepted

\bibitem[\protect\citeauthoryear{{Richardson}}{{Richardson}}{2007}]{Richardson%
_AGU07}
{Richardson}, J.~D. 2007, AGU Fall Meeting Abstracts, A3

\bibitem[\protect\citeauthoryear{{Richardson}, {Liu}, \& {Wang}}{{Richardson}
  et~al.}{2008}]{RLandW08}
{Richardson}, J.~D., {Liu}, Y.,  \& {Wang}, C. 2008, Advances in Space
  Research, 41, 237

\bibitem[\protect\citeauthoryear{{Richardson} et~al.}{{Richardson}
  et~al.}{1995}]{RPLandB95}
{Richardson}, J.~D., {Paularena}, K.~I., {Lazarus}, A.~J.,  \& {Belcher}, J.~W.
  1995, \grl, 22, 325

\bibitem[\protect\citeauthoryear{{Smith} et~al.}{{Smith}
  et~al.}{2006}]{SIMandR06}
{Smith}, C.~W., {Isenberg}, P.~A., {Matthaeus}, W.~H.,  \& {Richardson}, J.~D.
  2006, \apj, 638, 508

\bibitem[\protect\citeauthoryear{{Smith} et~al.}{{Smith}
  et~al.}{2001}]{SMZNOandR01}
{Smith}, C.~W., {Matthaeus}, W.~H., {Zank}, G.~P., {Ness}, N.~F., {Oughton},
  S.,  \& {Richardson}, J.~D. 2001, \jgr, 106, 8253

\bibitem[\protect\citeauthoryear{Sternal, Fichtner, \& Scherer}{Sternal
  et~al.}{2007}]{SFandS07}
Sternal, O., Fichtner, H.,  \& Scherer, K. 2007, Astron. Astrophys., to appear

\bibitem[\protect\citeauthoryear{{Stone} et~al.}{{Stone}
  et~al.}{2005}]{SCMcDHLandW05}
{Stone}, E.~C., {Cummings}, A.~C., {McDonald}, F.~B., {Heikkila}, B.~C., {Lal},
  N.,  \& {Webber}, W.~R. 2005, Science, 309, 2017

\bibitem[\protect\citeauthoryear{{Summers} \& {Thorne}}{{Summers} \&
  {Thorne}}{1991}]{SandT91}
{Summers}, D.,  \& {Thorne}, R.~M. 1991, Physics of Fluids B, 3, 1835

\bibitem[\protect\citeauthoryear{{Williams} \& {Zank}}{{Williams} \&
  {Zank}}{1994}]{WandZ94}
{Williams}, L.~L.,  \& {Zank}, G.~P. 1994, \jgr, 99, 19229

\bibitem[\protect\citeauthoryear{{Williams}, {Zank}, \& {Matthaeus}}{{Williams}
  et~al.}{1995}]{WZandM95}
{Williams}, L.~L., {Zank}, G.~P.,  \& {Matthaeus}, W.~H. 1995, \jgr, 100, 17059

\bibitem[\protect\citeauthoryear{Zank}{Zank}{1999}]{Zank99}
Zank, G.~P. 1999, Space Sci.\ Rev., 89, 413

\bibitem[\protect\citeauthoryear{{Zank}, {Matthaeus}, \& {Smith}}{{Zank}
  et~al.}{1996}]{ZMandS96}
{Zank}, G.~P., {Matthaeus}, W.~H.,  \& {Smith}, C.~W. 1996, \jgr, 101, 17093

\bibitem[\protect\citeauthoryear{Zank et~al.}{Zank et~al.}{1996}]{ZPWandH96}
Zank, G.~P., Pauls, H.~L., Williams, L.L.,  \& Hall, D.T. 1996, J. Geophys.
  Res., 101, 21639

\end{thebibliography}

\end{document}